\documentclass[a4paper,11pt]{article}
\pdfoutput=1 

\usepackage{jheppub} 

\usepackage[T1]{fontenc} 
\usepackage[english]{babel}
\usepackage{graphicx}			    	
\usepackage{slashed}
\usepackage{caption}
\usepackage{amsmath}
\usepackage{amsthm}
\usepackage{amsfonts}
\usepackage{mathtools}
\usepackage{subcaption}
\usepackage{tikz} 
\usepackage{tikz-feynman}
\usepackage{tikz-3dplot}
\tikzfeynmanset{compat=1.1.0}
\usepackage{placeins}
\usepackage{scalerel}
\usepackage{verbatim}
\usetikzlibrary{shapes.misc}
\tikzset{cross/.style={cross out, draw=black, minimum size=2*(#1-\pgflinewidth), inner sep=0pt, outer sep=0pt},
	cross/.default={5pt}}

\DeclareMathOperator{\Tr}{Tr}

\newcommand{\overbar}[1]{\mkern 1.5mu\overline{\mkern-1.5mu#1\mkern-1.5mu}\mkern 1.5mu}

\theoremstyle{plain}
\newtheorem{theorem}{Theorem}
\theoremstyle{plain}

\theoremstyle{plain}
\newtheorem{lemma}{Lemma}
\makeatletter
\def\@fpheader{\relax}
\makeatother

		\title{Nonresonant renormalization scheme for twist-$2$ operators in SU($N$) Yang-Mills theory}
		\author[b,a]{Francesco Scardino}
		
		\affiliation[a]{Physics Department, INFN Roma1, 
			Piazzale A. Moro 2, Roma, I-00185, Italy}
		\affiliation[b]{Physics Department, Sapienza University,Piazzale A. Moro 2, Roma, I-00185, Italy}
		
		\emailAdd{francesco.scardino@uniroma1.it}

		\abstract{			Recently, the short-distance asymptotics of the generating functional of  $n$-point correlators of twist-$2$ operators  in SU($N$) Yang-Mills (YM) theory has been worked out in \cite{BPSpaper2}. The above computation relies on a basis change of renormalized twist-$2$ operators, where  $-\gamma(g)/ \beta(g)$ reduces to $\gamma_0/ (\beta_0\,g)$ to all orders of perturbation theory, with $\gamma_0$ diagonal, $\gamma(g) = \gamma_0 g^2+\ldots$ the anomalous-dimension matrix  and $\beta(g) = -\beta_0 g^3+\ldots$ the beta function. The construction is based on a novel geometric interpretation of operator mixing \cite{Bochicchio:2021geometry}, under the assumption that the eigenvalues of the matrix $\gamma_0/ \beta_0$ satisfy the nonresonant condition $\lambda_i-\lambda_j\neq 2k$, with $\lambda_i$ in nonincreasing order and $k\in \mathbb{N}^+$. The nonresonant condition has been numerically verified up to $i,j=10^4$ in \cite{BPSpaper2}. In the present paper we provide a number theoretic proof of the nonresonant condition for twist-$2$ operators essentially based on the classic result that Harmonic numbers are not integers.
			Our proof in YM theory can be extended with minor modifications to twist-$2$ operators in $\mathcal{N}=1$ SUSY YM theory, large-$N$ QCD with massless quarks and massless QCD-like theories.}

\begin{document} 
			
			\definecolor{c969696}{RGB}{150,150,150}
			\maketitle	
			\flushbottom

\section{Introduction}
Recently, the ultraviolet (UV) asymptotics of  the generating functional of correlators of twist-$2$ operators in SU($N$) YM theory has been  explicitly calculated for the first time \cite{BPSpaper2}. It sets  strong UV constraints on the yet-to-come nonperturbative solution of large-$N$ YM theory and may be an essential guide for the search of such solution \cite{BPSpaper2}.\par
Besides, the above computation has also led to a refinement of the 't Hooft topological expansion in the large-$N$ SU($N$) theory \cite{BPSL} that is deeply related to the corresponding  nonperturbative effective theory of glueballs  \cite{BPSL}.\par
A crucial tool to perform the above calculation is a change of basis of renormalized twist-$2$ operators, where the renormalized mixing matrix $Z(\lambda)$ defined in Eq. \eqref{ZZ} becomes diagonal and one-loop exact to all orders of perturbation theory, based on a novel geometric interpretation of operator mixing \cite{Bochicchio:2021geometry} that we summarize as follows.\par
As recalled in the introduction of \cite{Becchetti:2021for}, in general, a change of renormalization scheme may consist both in a reparametrization of the coupling -- that changes the beta function $\beta(g) = -\beta_0 g^3+\ldots$, with $g\equiv g(\mu)$ the renormalized coupling -- and in a change of the basis of the operators that mix by renormalization  -- that changes the matrix of the anomalous dimensions $\gamma(g)=\gamma_0 g^2+ \cdots$. \par
In the present paper we are only interested in a change of the operator basis  \cite{Bochicchio:2021geometry}, while we keep the renormalization scheme for $\beta(g)$ fixed, for example the $\overbar{\text{MS}}$ scheme.\par
Of course, the above change of basis also affects the ratio $-\frac{\gamma(g)}{\beta(g)}$, with $\beta(g)$ fixed, in a way that we will explain below.\par
In the case of operator mixing, the renormalized Euclidean correlators
\begin{equation}\label{key}
	\langle \mathcal{O}_{k_1}(x_1)\ldots \mathcal{O}_{k_n}(x_n) \rangle = G^{(n)}_{k_1 \ldots k_n}( x_1,\ldots,  x_n; \mu, g(\mu))
\end{equation}
satisfy the Callan-Symanzik equation
\begin{align}\label{CallanSymanzik}
	& \Big(\sum_{\alpha = 1}^n x_\alpha \cdot \frac{\partial}{\partial x_\alpha} + \beta(g)\frac{\partial}{\partial g}+ \sum_{\alpha = 1}^n D_{\mathcal{O}_\alpha}\Big)G^{(n)}_{k_1 \ldots k_n} + \nonumber\\
	&+ \sum_a \Big(\gamma_{k_1a}(g) G^{(n)}_{ak_2 \ldots k_n}+ \gamma_{k_2a}(g) G^{(n)}_{k_1 a k_3 \ldots k_n} \cdots +\gamma_{k_n a}(g) G^{(n)}_{k_1 \ldots a}\Big) = 0\,,
\end{align}
with solution
\begin{align}\label{csformula}
	&G^{(n)}_{k_1 \ldots k_n}(\lambda x_1,\ldots, \lambda x_n; \mu, g(\mu)) \nonumber\\
	&= \sum_{j_1 \ldots j_n} Z_{k_1 j_1} (\lambda)\ldots Z_{k_n j_n}(\lambda)\hspace{0.1cm} \lambda^{-\sum_{i=1}^nD_{\mathcal{O}_{j_i}}} G^{(n)}_{j_1 \ldots j_n }( x_1, \ldots, x_n; \mu, g(\frac{\mu}{\lambda}))\,,
\end{align}
where $D_{\mathcal{O}_i}$ is the canonical dimension of $\mathcal{O}_i(x)$, with
\begin{equation}\label{eqZ}
	\Bigg(\frac{\partial}{\partial g} + \frac{\gamma(g)}{\beta(g)}\Bigg)Z(\lambda) = 0
\end{equation}
in matrix notation, and
\begin{equation} \label{ZZ}
	Z(\lambda) = P\exp \Big(\int_{g(\mu)}^{g(\frac{\mu}{\lambda})}\frac{\gamma(g')}{\beta(g')} dg'\Big)\,.
\end{equation}
We wonder whether a basis of renormalized operators exists where  $Z(\lambda)$ becomes diagonal, so that Eq. \eqref{csformula} greatly simplifies because it consists of only one term.\par
In a nutshell, in order to answer the above question, we interpret \cite{Bochicchio:2021geometry} a finite change of renormalization scheme -- i.e. a change of basis of the renormalized operators -- expressed in matrix notation
\begin{equation}\label{linearcomb}
	\mathcal{O}'(x) = S(g) \mathcal{O}(x)
\end{equation}
as a formal real-analytic invertible gauge transformation $S(g)$ \footnote{Obviously, in this context the gauge transformation $S(g)$ only depends on the coupling $g$ and it has nothing to do with the spacetime gauge group of the theory.} \cite{Bochicchio:2021geometry}. Under the action of $S(g)$, the matrix
\begin{equation}
	A(g) = -\frac{\gamma(g)}{\beta(g)} = \frac{1}{g} \Big(\frac{\gamma_0}{\beta_0} + \cdots\Big)
\end{equation}
associated to the differential equation for $Z(\lambda)$
\begin{equation}
	\Big(\frac{\partial}{\partial g} - A(g)\Big) Z(\lambda) = 0
\end{equation}
defines a connection $A(g)$
\begin{eqnarray} \label{sys2}
	A(g)= \frac{1}{g} \left(A_0 + \sum^{\infty}_ {n=1} A_{2n} g^{2n} \right)\,,
\end{eqnarray}
with a regular singularity at $g = 0$ that transforms as
\begin{equation}
	A'(g) = S(g)A(g)S^{-1}(g) + \frac{\partial S(g)}{\partial g} S^{-1}(g)\,,
\end{equation}
with
\begin{equation}
	\mathcal{D} = \frac{\partial }{\partial g} - A(g)
\end{equation}
the corresponding covariant derivative.
Consequently, $Z(\lambda)$ is interpreted as a Wilson line that transforms as
\begin{equation}
	Z'(\lambda) = S(g(\mu))Z(\lambda)S^{-1}(g(\frac{\mu}{\lambda}))\,.
\end{equation}
\begin{theorem}\label{th:1}  \cite{Bochicchio:2021geometry}
	If the matrix $\frac{\gamma_0}{\beta_0}$ is diagonalizable and nonresonant, i.e. its eigenvalues in nonincreasing order\\ $\lambda_1,\lambda_2,\ldots$ satisfy
	\begin{equation}
		\label{eq:nonresonant}
		\lambda_i-\lambda_j\neq 2k\ , \qquad i>j\ , \qquad k\in\mathbb{N}^+\,,
	\end{equation}
	then a formal holomorphic gauge transformation $S(g)$ exists that sets $A(g)$ in the canonical nonresonant form 
	\begin{equation} \label{1loop}
		A'(g) = \frac{\gamma_0}{\beta_0}\frac{1}{g}
	\end{equation}
	that is one-loop exact to all orders of perturbation theory. 
	As a consequence, $Z(\lambda)$ is diagonalizable as well, with eigenvalues
	\begin{equation}\label{eq:oneloop}
		Z_{\mathcal{O}_i}(\lambda) = \Bigg(\frac{g(\mu)}{g(\frac{\mu}{\lambda})}\Bigg)^{\frac{\gamma_{0\mathcal{O}_i}}{\beta_0}} \,,
	\end{equation}
	where $\gamma_{0\mathcal{O}_i}$ are the eigenvalues of $\gamma_0$.
\end{theorem}
In the present paper we demonstrate for twist-$2$ operators in SU($N$) YM theory that the matrix $\frac{\gamma_0}{\beta_0}$, which is known to be diagonal \cite{Belitsky:1998gc,Braun:2003rp}, satisfies the above nonresonant condition, thus proving the existence of the corresponding diagonal nonresonant renormalization scheme.  
\section{Plan of the paper}
In section \ref{sec:s1} we define the twist-$2$ operators in SU($N$) YM theory and report their one-loop anomalous dimensions \cite{Belitsky:2004sc}.\par
In section \ref{sec:s2} we recall some number-theoretic concepts, including the definition of $p$-adic order and the classic proof that the Harmonic numbers $H_n$ are not integers.  \par
In section \ref{sec:s3} we prove the nonresonant condition for twist-$2$ operators in SU($N$) YM theory.

\section{Anomalous dimensions of twist-$2$ operators in YM theory}\label{sec:s1}
\subsection{Twist-$2$ operators in YM theory}
The gauge-invariant collinear twist-$2$ operators in the light-cone gauge in the standard basis read \cite{BPSpaper2,BPS1}
\begin{align} \label{1000}
	\nonumber
	\mathbb{O}_{s} = &\Tr \partial_{+} \bar{A}(x)(i\overrightarrow{\partial}_++i\overleftarrow{\partial}_+)^{s-2} C^{\frac{5}{2}}_{s-2}\left(\frac{\overrightarrow{\partial}_+-\overleftarrow{\partial}_+}{\overrightarrow{\partial}_++\overleftarrow{\partial}_+}\right)\partial_{+} {A}(x)\nonumber\\
	\tilde{\mathbb{O}}_{s}
	= &\Tr \partial_{+} \bar{A}(x)(i\overrightarrow{\partial}_++i\overleftarrow{\partial}_+)^{s-2} C^{\frac{5}{2}}_{s-2}\left(\frac{\overrightarrow{\partial}_+-\overleftarrow{\partial}_+}{\overrightarrow{\partial}_++\overleftarrow{\partial}_+}\right)\partial_{+} {A}(x)\nonumber \\
	\mathbb{S}_{s}
	=&\frac{1}{\sqrt{2}}\Tr \partial_{+} \bar{A}(x)(i\overrightarrow{\partial}_++i\overleftarrow{\partial}_+)^{s-2} C^{\frac{5}{2}}_{s-2}\left(\frac{\overrightarrow{\partial}_+-\overleftarrow{\partial}_+}{\overrightarrow{\partial}_++\overleftarrow{\partial}_+}\right)\partial_{+} \bar{A}(x)\nonumber\\
	\bar{\mathbb{S}}_{s} =&\frac{1}{\sqrt{2}}\Tr \partial_{+} A(x)(i\overrightarrow{\partial}_++i\overleftarrow{\partial}_+)^{s-2} C^{\frac{5}{2}}_{s-2}\left(\frac{\overrightarrow{\partial}_+-\overleftarrow{\partial}_+}{\overrightarrow{\partial}_++\overleftarrow{\partial}_+}\right)\partial_{+} A(x)\,,
\end{align}
where $C^{\alpha'}_l(x)$ are Gegenbauer polynomials that are a special case of  Jacobi polynomials \cite{BPS1}.\par
These operators are the restriction, up to perhaps normalization and linear combinations \cite{BPS1}, to the component with maximal-spin projection $s$ along the $p_+$ direction of the balanced,
$\mathbb{O}^{\mathcal{T}=2}_{s}, \tilde{\mathbb{O}}^{\mathcal{T}=2}_{s}$, and unbalanced, $\mathbb{S}^{\mathcal{T}=2}_{s}$, twist-$2$ operators that to the leading and next-to-leading order of perturbation theory transform as primary operators with respect to the conformal group \cite{Beisert:2004fv}
\begin{align} \label{1}
	&\mathbb{O}^{\mathcal{T}=2}_{\rho_1\ldots \rho_s}= \Tr\, F^\mu_{(\rho_1}\overleftrightarrow{D}_{\rho_2}\ldots \overleftrightarrow{D}_{\rho_{s-1}}F_{\rho_s)\mu}-\,\text{traces}\qquad\qquad \nonumber\\
	&\tilde{\mathbb{O}}^{\mathcal{T}=2}_{\rho_1\ldots \rho_s}= \Tr\, \tilde{F}^\mu_{(\rho_1}\overleftrightarrow{D}_{\rho_2}\ldots \overleftrightarrow{D}_{\rho_{s-1}}F_{\rho_s)\mu}-\,\text{traces}\qquad\qquad \nonumber\\
	&\mathbb{S}^{\mathcal{T}=2}_{\mu\nu\rho_1\ldots \rho_{s-2}\lambda\sigma} \,\,= \Tr\, (F_{\mu(\nu}+i\tilde{F}_{\mu(\nu})\overleftrightarrow{D}_{\rho_1}\ldots \overleftrightarrow{D}_{\rho_{s-2}}(F_{\lambda)\sigma}+i\tilde{F}_{\lambda)\sigma})-\,\text{traces} \nonumber\\
	&\bar{\mathbb{S}}^{\mathcal{T}=2}_{\mu\nu\rho_1\ldots \rho_{s-2}\lambda\sigma} \,\,= \Tr\, (F_{\mu(\nu}-i\tilde{F}_{\mu(\nu})\overleftrightarrow{D}_{\rho_1}\ldots \overleftrightarrow{D}_{\rho_{s-2}}(F_{\lambda)\sigma}-i\tilde{F}_{\lambda)\sigma})-\,\text{traces}\,,
\end{align}
where the parentheses stand for symmetrization of all the indices in between and the subtraction of the traces ensures that the contraction of any two indices is zero.
\subsection{Anomalous dimensions}\label{sec:s12}

The maximal-spin components of the above operators $\mathcal{O}_s$ only mix with the derivatives along the direction $p_+$ of the same operators with lower spin and same canonical dimensions \cite{Braun:2003rp,Belitsky:1998gc}. We define the bare operators with $s\geq k$ as
\begin{equation}
	\mathcal{O}^{B\,(k)}_s =(i\partial_+)^{k}\mathcal{O}^{B}_s
\end{equation}
that, to the leading order of perturbation theory, for $k >0$ are conformal descendants of the corresponding primary conformal operator $\mathcal{O}^{B\,(0)}_s = \mathcal{O}^{B}_s $. 
As a consequence of operator mixing, we obtain for the renormalized operators \cite{Braun:2003rp,Belitsky:1998gc}
\begin{equation}\label{m}
	\mathcal{O}^{(k)}_s= \sum_{s\geq i \geq 2} Z_{si}\, \mathcal{O}^{B\,(k+s-i)}_{i}\,,
\end{equation}
where the bare mixing matrix $Z$ is lower triangular\footnote{$Z$, which in dimensional regularization on depends on $g$ and $\epsilon$, should not be confused with $Z(\lambda)$.} \cite{Braun:2003rp,Belitsky:1998gc}.\par
Hence, in general the anomalous-dimension matrix $\gamma(g)$ is lower triangular, but with  $\gamma_0$  diagonal. 
In our notation the eigenvalues of $\gamma_0$  are given \cite{Belitsky:1998gc} by 
\begin{align}
	\label{an1}
	\gamma^{\mathbb{O}}_{0s}= \frac{2}{(4 \pi)^2} \Big(4 \psi(s+1) - 4 \psi(1) -\frac{11}{3}- 8 \frac{s^2+s+1}{(s-1)s(s+1)(s+2)}\Big)
\end{align}
for  $\mathbb{O}_{s}$ with even $s\geq 2$, and
\begin{align}
	\label{an2}
	\gamma^{\mathbb{\tilde{O}}}_{0s}= \frac{2}{(4 \pi)^2}  \Big(4 \psi(s+1) - 4 \psi(1) -\frac{11}{3} - 8 \frac{s^2+s-2}{(s-1)s(s+1)(s+2)}  \Big)
\end{align}
for $\tilde{\mathbb{O}}_{s}$ with odd $s\geq 3$, where  $\psi(i)$ is the digamma function. Consistently, $\gamma_{02}=0$ \footnote{The anomalous dimension of the stress-energy tensor is zero.}. 
The eigenvalues of $\gamma_0$ for $\mathbb{S}_{s}$ and $\bar{\mathbb{S}}_{s}$ are given by  \cite{Belitsky:2004sc}
\begin{equation}
	\gamma^{\mathbb{S}}_{0s}=  \frac{2}{(4 \pi)^2} \left(4\psi(s+1)-4\psi(1)-\frac{11}{3}\right)\,,
\end{equation}
with even $s\geq 2$.\par
Hence, the eigenvalues of $\gamma_0$ are naturally increasingly ordered for increasing $s$, contrary to the ordering in theorem \ref{th:1}. Yet, it is easily seen from the proof \cite{Bochicchio:2021geometry} of theorem \ref{th:1} that in this case the nonresonant condition becomes
\begin{equation}
	\label{eq:nonresonant2}
	\lambda_j-\lambda_i\neq 2k\ , \qquad j>i\ , \qquad k\in\mathbb{N}^+\,,
\end{equation}
with $\lambda_1\leq \lambda_2\leq \lambda_3\leq \ldots$.
\section{Number-theoretic concepts}\label{sec:s2}
\subsection{$p$-adic order}
We define the $p$-adic  order of an integer $n$ as the exponent of the highest power of the prime number $p$ that divides $n$ \cite{ireland1990classical}. 
Namely, the $p$-adic order of an integer is the function
\begin{equation}
	\nu_p(n)=
	\begin{cases}
		\mathrm{max}\{k \in \mathbb{N} : p^k\,\text{divides}\,n\} & \text{if } n \neq 0\\
		\infty & \text{if } n=0\,.
	\end{cases}
\end{equation}
For instance, $\nu_3(24) = \nu_3(3\times2^3) = 1$ and $\nu_2(24) = 3$.\par
The $p$-adic order can be extended to rational numbers by the property \cite{ireland1990classical}
\begin{equation}
	\label{eq:p1}
	\nu_p\left(\frac{a}{b}\right)  =\nu_p(a)-\nu_p(b)\,.
\end{equation}
Therefore, rational numbers can have negative $p$-adic order, while integer numbers can only have positive or zero values for all $p$.
Other properties are \cite{ireland1990classical}
\begin{align}
	\label{eq:p2}
	&\nu_p(a\cdot b) = \nu_p(a)+\nu_p(b)\nonumber\\
	&\nu_p(a+b)\geq\min\bigl\{ \nu_p(a), \nu_p(b)\bigr\}\,.
\end{align}
If  $\nu_p(a) \neq \nu_p(b)$ 
\begin{equation}
	\label{equal}
	\nu_p(a+b)= \min\bigl\{ \nu_p(a), \nu_p(b)\bigr\}
\end{equation}
\cite{ireland1990classical} that is crucial for the proof below.
\subsection{Harmonic numbers and Bertrand's postulate}
Bertrand's postulate\footnote{It is actually a theorem.}  \cite{cebby,ramanujan1919proof} states that for every real  $x\geq2$ there  exists at least one prime number $p$ satisfying
\begin{equation}\label{eq:bertrand}
	\frac{x}{2}+1\leq p\leq x\,.
\end{equation} 
This means that  for every $p\in\left[\frac{x}{2}+1,x\right]$ its double, $2p$, cannot be in the same interval because $2p\geq x+2$. \par
We  apply Bertrand's postulate to demonstrate the classic result that the Harmonic numbers $H_n$ 
\begin{align}\label{eq:Hn}
	H_n = \sum_{k = 1}^{n}\frac{1}{k}
\end{align}
are not integers for all $n\geq 2$.
\subsection{Standard argument}
Let  $p$ be a prime in the interval 
\begin{equation}\label{eq:postulaten}
	\frac{n}{2}+1\leq p\leq n\,.
\end{equation}
For such $p$, $\frac{1}{p}$ appears in the sum of Eq. \eqref{eq:Hn}, but there is no $k>p$ that can have $p$ as a prime factor because its prime factor decomposition should at least contain $2p$ that is outside the above interval. Therefore, except for $\frac{1}{p}$, every  term $\frac{1}{k}$ in the sum has $k$ divisible only by primes different than $p$. As a consequence, if we set
\begin{equation}
	\sum_{k=1}^n\frac{1}{k} = \frac{1}{p} + \frac{a}{b}\,,
\end{equation}
then $b$ is not divisible by $p$, i.e. $\gcd(b,p)=1$. This implies that Harmonic numbers are not integers. Indeed,
\begin{equation}
	\nu_p\left(\frac{1}{p}\right)=-1\qquad \nu_p\left(\frac{a}{b}\right) = \nu_p(a)>0\,.
\end{equation} 
Hence, by noticing that  $\nu_p\left(\frac{1}{p}\right)\neq \nu_p\left(\frac{a}{b}\right)$, it follows from Eq. \eqref{equal}
\begin{equation}
	\nu_p(\frac{1}{p}+\frac{a}{b}) =\min(\nu_p\left(\frac{1}{p}\right),\nu_p\left(\frac{a}{b}\right)) = -1
\end{equation}
and, finally,
\begin{equation}
	\nu_p(H_n) = \nu_p(\frac{1}{p} + \frac{a}{b}) = \min(\nu_p(\frac{1}{p}),\nu_p(\frac{a}{b})) = -1\,.
\end{equation}
Therefore, since the $p$-adic order of $H_n$ is negative, it cannot be an integer. \\
We will refer to this argument as the \textit{standard} one, since it will be repeatedly applied in the following. 
\subsection{Generalized argument}\label{sec:generalizedsec}
We now study more complicated sums than Harmonic numbers
\begin{equation}\label{eq:general}
	K_n = \sum_{k=1}^{n}\frac{c_k}{k}\,.
\end{equation}
We begin considering the case where the coefficients $c_k$ can take positive and negative values $\pm1$  or $\pm2$.\par
Clearly, the coefficients $\pm1$ does not alter the standard argument, as $1$ is coprime with any $p$ and, if $p$ has been found by Bertrand's postulate, there is no $k\neq p$ in the sum that has $p$ in its prime factorization. Thus, $K_n$ is not integer by the standard argument. \par 
A bit more care is needed if $\pm2$ occurs in the numerators.  If $n\geq3$, by Betrand's postulate, it exists a prime $ p\geq 3$ in Eq. \eqref{eq:postulaten} such that 
\begin{equation}
	K_n  = \frac{c_p}{p}+\frac{a}{b}
\end{equation}
where $\gcd(b,p)=\gcd(c_p,p)=1$. Indeed, $c_p$ and $p$ are coprime in this case and, as above, even if the other $c_k$ could take the values $\pm1,\pm2$, there is no $k$ in the sum that has $p$ as a prime factor. \par 
Indeed, let us suppose there are 
\begin{equation}\label{eq:pp}
	k_1=p'
\end{equation}
and
\begin{equation}\label{eq:2pp}
	k_2=2p'
\end{equation}
such that $c_{k_2}=\pm2$, then the terms $\frac{c_{k_1}}{k_1}=\frac{c_{k_1}}{p'}$ and $\frac{\pm 2}{k_2}=\frac{\pm1}{p'}$ combine together as $\frac{c_{p'}}{p'}$, where $c_{p'}$ could be zero. However, $p'$ by eqs. \eqref{eq:pp} and \eqref{eq:2pp} is not one of the primes selected by Bertrand's postulate, so that its possible absence does not affect the standard argument, which still applies.\par
More generally, if $c_k\in \mathbb{Z} \setminus \{0\}$ and one can find a prime $p$ satisfying Eq. \eqref{eq:postulaten} such that $\gcd(c_p,p)=1$, then the standard argument still applies. Indeed, by Bertrand's postulate, the terms in the sum different from $\frac{c_p}{p}$ combine in a fraction whose denominator does not contain $p$ in its prime factorization.\par 
We consider now the last and most complicated case, where we allow some $c_k$ to be zero. In this case, there must exist at least one prime $p$ satisfying Eq. \eqref{eq:postulaten} such that $c_p\neq 0$ and $\gcd(c_p,p)=1$. If this condition is satisfied, then there can be an arbitrary number of zero values of $c_k$ and the standard argument still applies to prove that $K_n$ is not an integer.\par 
Clearly, the above condition requires a direct inspection of the sum on a case by case basis.

\section{Proof of the nonresonant condition}\label{sec:s3}
In this section we  prove  that the eigenvalues of the anomalous dimensions of the above twist-$2$ operators are nonresonant according to Eq. \eqref{eq:nonresonant2}.  \par
We recall that the digamma function can be written as
\begin{equation}
	\label{gammaH}
	\psi(n+1) = H_n-\gamma\,,
\end{equation}
where $\gamma$ is the Euler-Mascheroni constant. 
\subsection{Nonresonant condition for unbalanced twist-$2$ operators}\label{sec:proof1}
Using Eq. \eqref{gammaH}, we write in a more suitable form the anomalous dimension of $\mathbb{S}_s$
\begin{align}
	\gamma^{\mathbb{S}}_{0n}=\frac{2}{(4 \pi)^2} \left(4H_n-\frac{11}{3}\right)\,,
\end{align}
with $n=2,4,6,\ldots$.
\begin{lemma}\label{lemma:1}
	The sequence $\gamma^{\mathbb{S}}_{0n}$ is  monotonically increasing
	\begin{equation}
		\gamma^{\mathbb{S}}_{0n+1}\geq \gamma^{\mathbb{S}}_{0n}
	\end{equation}
\end{lemma}
Proof.\\\\
\begin{equation}
	\gamma^{\mathbb{S}}_{0n+1}-\gamma^{\mathbb{S}}_{0n} = \frac{8}{(4\pi)^2}\frac{1}{n+1}> 0 \,.
\end{equation}
\qed\\\\
Therefore, the sequence  $\gamma_{0n}^{\mathbb{S}}$ is increasing and matches the ordering in Eq. \eqref{eq:nonresonant2}.\\
\begin{theorem}
	The eigenvalues of $\frac{\gamma^{\mathbb{S}}_0}{\beta_0}$ are nonresonant
	\begin{equation}\label{eq:th1}
		\frac{\gamma_{0n}^{\mathbb{S}}-\gamma_{0m}^{\mathbb{S}}}{\beta_0} \neq 2k\,,\qquad k\in\mathbb{N}^+,\quad \forall n>m\geq2\,,
	\end{equation}
	where $\beta_0 = \frac{1}{(4\pi)^2}\frac{11}{3}$.\\ 
\end{theorem}
Proof.\\ \\
We choose $n =m+x$, where $x\geq1$ is a natural number. Then Eq. \eqref{eq:th1} reads
\begin{equation}
	\Delta_m^{\mathbb{S}} (x)=\frac{\gamma_{0{m+x}}^{\mathbb{S}}-\gamma_{0m}^{\mathbb{S}}}{\beta_0} = \frac{24}{11}\sum_{k = m+1}^{m+x}\frac{1}{k} = \frac{24}{11}\Sigma_m(x)\,,
\end{equation}
with
\begin{equation}
	\Sigma_m(x) = \sum_{k = m+1}^{m+x}\frac{1}{k}\,.
\end{equation}
We  parametrize $x$ as
\begin{align}
	x = m+t\qquad t\geq0\,,
\end{align}
that leaves out of the first part of the proof all possible values of $x<m$, which we will consider later. Hence,
\begin{align}
	\Delta_m^{\mathbb{S}} (m+t)= \frac{24}{11}\sum_{k = m+1}^{2m+t}\frac{1}{k}\,.
\end{align}
Again, by Bertrand's postulate  it exists a prime between 
\begin{equation}\label{eq:pin1}
	m+1+\frac{t}{2}\leq p\leq 2m+t\,.
\end{equation}
Hence, by the standard argument 
\begin{equation}
	\nu_p(\Sigma_m(m+t)) =-1\,.
\end{equation}
Therefore, by using eqs. \eqref{eq:p1} and \eqref{eq:p2}
\begin{align}
	\nu_p(\frac{24}{11}	\Sigma_m(m+t)) &= \nu_p(\frac{24}{11})+\nu_p(	\Sigma_m(m+t)) \nonumber\\
	&= \nu_p(24)-\nu_p(11)-1\,.
\end{align} 
Then, if $m>2$ or $m=2$ and $t>0$,  by Eq. \eqref{eq:pin1} $p\geq 5$, so that $\nu_p(24)=0$ and
\begin{equation}
	\nu_p(\frac{24}{11}	\Sigma_m(m+t)) <0.
\end{equation}
Besides, for the special case  $m=2$ and $t= 0$, $\Sigma_2(2) = \frac{7}{12}$ and  $	\Delta_2^{\mathbb{S}} (2) = \frac{14}{11}$. We conclude that, if $x\geq m$, the $p$-adic order of 
\begin{equation}
	\nu_p(\Delta_m^{\mathbb{S}}(x))<0
\end{equation}
and therefore it cannot be an integer.\par
In the proof above we have left out the values of $x< m$.
First, we demonstrate that
\begin{align}
	\label{em1}
	\Sigma_m(x) <\log(2)<1\qquad\qquad\forall x<m\,.
\end{align}	
Indeed,
\begin{equation}\label{eq:ineq}
	\Sigma_m(x)\leq\Sigma_m( m-1) <	\lim_{m\rightarrow\infty}\Sigma_m( m-1)  = \log(2)
\end{equation}
because, by Lemma \ref{lemma:1}, $\Sigma_m(x)$ is monotonically increasing in $x$ for fixed $m$ and $\Sigma_m(m-1)$ is monotonically increasing.\par
The limit in Eq. \eqref{eq:ineq} is computed by means of the Euler-Maclaurin formula \cite{EMformula}
\begin{equation}
	\Sigma_m( m-1) = \log(2)-\frac{3}{4 m}+\frac{1}{16 m^2}+O\left(\frac{1}{m^4}\right)\,.
\end{equation}
Hence, Eq. \eqref{eq:ineq} implies
\begin{align}
	\Delta_m^{\mathbb{S}}(x)= 	\frac{24}{11}\Sigma_m(x) <\frac{24}{11}\log(2)<1.53\qquad \forall x<m\, .
\end{align}
Therefore, $\Delta_m^{\mathbb{S}}(x)$  is strictly lower than $2$.\\
We conclude that $\Delta_m^{\mathbb{S}}$ cannot be an integer for $x\geq m$ and cannot be an integer larger then $1$ for $x<m$, hence the theorem is proved.
\qed
\subsection{Nonresonant condition for balanced twist-$2$ operators of even spin}
We now study the anomalous dimension of balanced operators of even spin
\begin{equation}
	\gamma^{\mathbb{O}}_{0n}= \frac{2}{(4 \pi)^2} \left(4 H_n-\frac{11}{3} - 8 \frac{n^2+n+1}{(n-1)n(n+1)(n+2)}\right)
\end{equation}
with $n=2,4,6,\ldots$.
\begin{lemma}\label{lemma2}
	The sequence $\gamma^{\mathbb{O}}_{0n}$ is  monotonically increasing
	\begin{equation}
		\gamma^{\mathbb{O}}_{0n+1}\geq \gamma^{\mathbb{O}}_{0n}
	\end{equation}
\end{lemma}
Proof.\\\\
We write the difference of consecutive eigenvalues as
\begin{align}
	\gamma^{\mathbb{O}}_{0n+1}-\gamma^{\mathbb{O}}_{0n}	& = \frac{8}{(4\pi)^2}\Bigg(\frac{1}{n+1}-2\frac{(n+1)^2+n+2}{n(n+1)(n+2)(n+3)}+2\frac{n^2+n+1}{(n-1)n(n+1)(n+2)}\Bigg)\nonumber\\
	&= \frac{8}{(4\pi)^2}\left(-\frac{2}{n}+\frac{3}{n+1}-\frac{2}{n+2}+\frac{1}{n+3}+\frac{1}{n-1}\right)\,.
\end{align}
We collect as follows the terms inside the parentheses
\begin{align}
	&\frac{1}{n-1}-\frac{1}{n}\nonumber\\
	+&\frac{2}{n+1}-\frac{2}{n+2}\nonumber\\
	+&\frac{1}{n+1}+\frac{1}{n+3}-\frac{1}{n}\,,
\end{align}
so that the first two lines are obviously positive. The third line is also positive, since for $n\geq 2$ 
\begin{align}
	\frac{1}{n+1}+\frac{1}{n+3}-\frac{1}{n} = \frac{n^2-3}{n (n+1) (n+3)}>0\,.
\end{align}
Therefore, $\gamma^{\mathbb{O}}_{0n}$ is monotonically increasing and matches the ordering in Eq. \eqref{eq:nonresonant2}.
\qed\\

\begin{theorem}\label{th:2}
	The eigenvalues of $\frac{\gamma^{\mathbb{O}}_0}{\beta_0}$ are nonresonant
	\begin{equation}
		\frac{\gamma_{0n}^{\mathbb{O}}-\gamma_{0m}^{\mathbb{O}}}{\beta_0} \neq 2k\,,\qquad k\in\mathbb{N}^+,\quad \forall n>m\geq 2\,,
	\end{equation}
	where $\beta_0 = \frac{1}{(4\pi)^2}\frac{11}{3}$.
\end{theorem} 
Proof.\\ \\
The proof is very similar to the one above for the unbalanced twist-$2$ operators. However, as outlined in section \ref{sec:generalizedsec}, some extra care is needed.\par
We choose again $n =m+x$, where $x>0$ is a natural number. Then, the difference of eigenvalues can be written as
\begin{align}
	\Delta^{\mathbb{O}} _m(x)&=\frac{\gamma_{0{m+x}}^{\mathbb{O}}-\gamma_{0m}^{\mathbb{O}}}{\beta_0}\nonumber\\
	&= \frac{24}{11}\Bigg(\sum_{k = m+1}^{m+x}\frac{1}{k}+2\frac{m^2+m+1}{(m-1)m(m+1)(m+2)}\nonumber\\
	&\quad-2\frac{(m+x)^2+m+x+1}{(m+x-1)(m+x)(m+x+1)(m+x+2)}\Bigg)\,,
\end{align}
that is simplified as follows
{\small
	\begin{align}
		\Delta_m^{\mathbb{O}} (x)= &\frac{24}{11}\Big(\frac{1}{m-1}-\frac{1}{m}+\frac{1}{m+1}-\frac{1}{m+2}+\sum_{k = m+1}^{m+x}\frac{1}{k}\nonumber\\
		&-\frac{1}{m+x-1}+\frac{1}{m+x}-\frac{1}{m+x+1}+\frac{1}{m+x+2}\Big)\nonumber\\
		=&\frac{24}{11}K_m(x)\,,
	\end{align}
}
with
{\small
	\begin{align}\label{eq:si}
		K_m(x) =&\frac{1}{m-1}-\frac{1}{m}+\frac{1}{m+1}-\frac{1}{m+2}+\sum_{k = m+1}^{m+x}\frac{1}{k}\nonumber\\
		&-\frac{1}{m+x-1}+\frac{1}{m+x}-\frac{1}{m+x+1}+\frac{1}{m+x+2}\,.
	\end{align}
}
This is a sum, with possibly alternating signs,  that goes from $m-1$ to $m+x+2$, with a gap at $\frac{1}{m+2}$ and $\frac{1}{m+x-1}$. We rewrite the sum as 
{\small
	\begin{align}\label{eq:si2}
		K_m(x) =&\frac{1}{m-1}-\frac{1}{m}+\frac{2}{m+1}+\sum_{k = m+3}^{m+x-2}\frac{1}{k}+\frac{2}{m+x}-\frac{1}{m+x+1}+\frac{1}{m+x+2}\,.
	\end{align}
}

The above sum is of the kind of Eq. \eqref{eq:general} with $c_k=0,\pm1,\pm2$, where specifically only two values of $c_k$ are actually zero, $c_{m+2}=c_{m+x-1}=0$. According to the generalized argument, the presence of gaps in the sum requires a case by case inspection.\par
As before, we  parametrize $x$ 
\begin{align}
	x = m+t\qquad t\geq0\,,
\end{align}
that leaves out of Eq. \eqref{eq:si} a finite number of possible values of $x$, namely $x<m$, which we will address later. Therefore, for $x\geq m$ 
\begin{align}
	\Delta_m^{\mathbb{O}} (m+t)&=\frac{24}{11}\Big(\frac{1}{m-1}-\frac{1}{m}+\frac{2}{m+1}+\sum_{k = m+3}^{2m+t-2}\frac{1}{k}\nonumber\\
	&\quad+\frac{2}{2m+t}-\frac{1}{2m+t+1}+\frac{1}{2m+t+2}\Big)\,,
\end{align}
the corresponding gaps occurring for $k=m+2$ and $k=2m+t-1$.\par
We now use Bertrand's postulate that there exists a prime $p$ between 
\begin{equation}
	m+\frac{t}{2}+2\leq p\leq2m+t+2\,.
\end{equation}
Hence, with the extra assumption that $p\neq 2m+t-1$ and $p \neq m+2$,  by the generalized argument
\begin{equation}\label{eq:res2}
	\nu_p(\Delta_m^{\mathbb{O}} (x))<0
\end{equation}
for $x\geq m$, so  that $\Delta_m^{\mathbb{O}} (x)$ is not an integer.\par
Let us suppose now that $p = m+2$ is the only prime in the interval $m+\frac{t}{2}+2\leq p\leq2m+t+2$. This could happen only for $t=0$, Eq. \eqref{eq:res2} still holding for all  $t>0$.\par 
Let $p_n$ be the $n$-th prime.  The "gap between consecutive primes" function $g(p_n)$, defined by
\begin{equation}
	p_{n+1} = p_n+g(p_n)+1
\end{equation}
satisfies the inequality \cite{Negura} 
\begin{equation}
	g(p_n)<\frac{1}{5}p_n\qquad\qquad\forall p>23\,.
\end{equation}
We  relax the above inequality to make it work for all primes 
\begin{equation}
	g(p_n)\leq\frac{1}{2}p_n\qquad\qquad\forall p\,.
\end{equation}
Therefore,
\begin{equation}\label{eq:pn1}
	p_n<p_{n+1} \leq \frac{3}{2}p_n+1\,.
\end{equation}
By coming back to our proof, if $p_n = m+2$ is the only prime satisfying
\begin{equation}\label{eq:pm}
	m+2\leq p\leq2m+2\,,
\end{equation}
then the next prime $p_{n+1}$ must be outside the interval of Eq. \eqref{eq:pm}, so that by Eq. \eqref{eq:pn1}  
\begin{equation}\label{eq:absurd}
	2m+2<p_{n+1}\leq\frac{3}{2}p_n+1=\frac{3}{2}(m+2)+1\,.
\end{equation}
Eq. \eqref{eq:absurd} can only be true if $m<4$. Hence, if $m\geq 4$, $p_n=m+2$ is not the the only prime satisfying Eq. \eqref{eq:pm} because $p_{n+1}$ also satisfies Eq. \eqref{eq:pm}, that is a contradiction. Hence, only for $m=2,3$  $p_n=m+2$ can be the only prime in the interval. As a consequence, we  verify by direct computation that $\Delta_2^{\mathbb{O}}(2)$ and $\Delta_3^{\mathbb{O}}(3)$ are not an integers.\par

Next, we consider the other gap in the sum, if $p_n=2m+t-1$ is the only prime satisfying 
\begin{equation}\label{eq:interval2}
	m+2+\frac{t}{2}\leq p\leq 2m+2+t\,.
\end{equation}
Then, we look for the location of the previous prime $p_{n-1}$. By Eq. \eqref{eq:pn1} it follows 
\begin{equation}
	\frac{2}{3}(p_n-1)\leq p_{n-1}<p_n\,,
\end{equation}
that for $p_n = 2m+t-1$ reads
\begin{equation}\label{eq:ineqnm1}
	\frac{2}{3}(2m+t-2)\leq p_{n-1}\leq 2m+t-2\,.
\end{equation}
The above equation implies that $p_{n-1}$ is inside the interval in Eq. \eqref{eq:interval2}  
\begin{equation}\label{eq:mt1}
	p_{n-1}\geq	\frac{2}{3}(2m+t-2)\geq m+2+\frac{t}{2}
\end{equation}
provided that  
\begin{equation}
	m+\frac{t}{2}\geq 10
\end{equation}
that contradicts the assumption that $p_n$ is the only such prime. \par
Hence, the only values of $m$ and $t$ for which a $p_n=2m+t-1$ is the only prime satisfying Eq. \eqref{eq:interval2} are
\begin{equation}
	m+\frac{t}{2}< 10\,.
\end{equation}
We have checked by direct computation that the corresponding $\Delta_m^{\mathbb{O}}(m+t)$ are not integers.
This concludes the examination of the exceptional values corresponding to the gaps, according to the generalized argument.\par
In the proof above  we have  left out the values of $x< m$.
In this case, in analogy with section \ref{sec:proof1}, we demonstrate the bound
\begin{align}
	\label{em}
	K_m(x)<\log(2)<1\,,\qquad \qquad\forall x<m\,.
\end{align}	
Indeed,
\begin{equation}\label{eq:ineq2}
	K_m(x)\leq K_m( m-1) <	\lim_{m\rightarrow\infty}K_m( m-1)  = \log(2)
\end{equation}
because $K_m(x)$ is monotonically increasing in $x$ for fixed $m$ and $K_m(m-1)$ is monotonically increasing in $m$ by Lemma \ref{lemma2}.\par
The limit in Eq. \eqref{eq:ineq2} is calculated by means of the Euler-Maclaurin formula \cite{EMformula}
\begin{equation}
	K_m( m-1) =\log(2)-\frac{3}{4 m}+\frac{25}{16 m^2}-\frac{9}{4 m^3}+O\left(\frac{1}{m^4}\right)\,.
\end{equation}
Hence, by Eq. \eqref{eq:ineq2}
\begin{align}
	\Delta_m^{\mathbb{O}}(x)= 	\frac{24}{11}K_m(x) <\frac{24}{11}\log(2)<1.53\,\qquad\forall x<m\,.
\end{align}
Therefore, if $\Delta_m^{\mathbb{O}}(x)$ is to be an integer, it could only be $1$.\par
We conclude that $\Delta_m^{\mathbb{O}}$ cannot be an integer for $x\geq m$ and cannot be an integer larger then $1$ for $x<m$, thus proving the theorem.
\qed
\subsection{Nonresonant condition for balanced twist-$2$ operators of odd spin}
We are left with the  anomalous dimension of the odd spin balanced operators.
{\small
	\begin{align}
		\gamma^{\tilde{\mathbb{O}}}_{0n}
		&= \frac{2}{(4 \pi)^2}  \left(4 \psi(n+1) - 4 \psi(1) -\frac{11}{3} - 8 \frac{n^2+n-2}{(n-1)n(n+1)(n+2)}  \right)\nonumber\\
		&= \frac{2}{(4 \pi)^2}  \left(4 H_n-\frac{11}{3} - 8 \frac{n^2+n-2}{(n-1)n(n+1)(n+2)} \right)
	\end{align}
}
with $n=3,5,7,\ldots$.
\begin{lemma}\label{lemma3}
	The sequence $\gamma^{\tilde{\mathbb{O}}}_{0n}$ is  monotonically increasing
	\begin{equation}
		\gamma^{\tilde{\mathbb{O}}}_{0n+1}\geq \gamma^{\tilde{\mathbb{O}}}_{0n}
	\end{equation}
\end{lemma}
Proof.\\\\
We explicitly write the difference
\begin{align}
	\gamma^{\tilde{\mathbb{O}}}_{0n+1}- \gamma^{\tilde{\mathbb{O}}}_{0n}
	& = \frac{8}{(4\pi)^2}\Bigg(\frac{1}{n+1}-2\frac{(n+1)^2+n-1}{n(n+1)(n+2)(n+3)}+2\frac{n^2+n-2}{(n-1)n(n+1)(n+2)}\Bigg)\nonumber\\
	&= \frac{8}{(4\pi)^2}\frac{n^2+2 n+4}{n^3+3 n^2+2 n} >0\,.
\end{align}
Therefore, $\gamma^{\tilde{\mathbb{O}}}_{0n}$ is a monotonically increasing and matches the ordering in Eq. \eqref{eq:nonresonant2}.\\
\qed\\\\

\begin{theorem}
	The eigenvalues of $\frac{\gamma^{\tilde{\mathbb{O}}}_0}{\beta_0}$ are nonresonant
	\begin{equation}
		\frac{\gamma_{0n}^{\tilde{\mathbb{O}}}-\gamma_{0m}^{\tilde{\mathbb{O}}}}{\beta_0} \neq 2k\,,\qquad k\in\mathbb{N}^+,\quad \forall n>m\geq 3\,,
	\end{equation}
	where $\beta_0 = \frac{1}{(4\pi)^2}\frac{11}{3}$.
\end{theorem} 
Proof.\\\\ 
We choose $n =m+x$, where $x>0$ is a natural number. Then, the difference of eigenvalues reads
\begin{align}
	\Delta_m^{\tilde{\mathbb{O}}}(x)&=\frac{\gamma_{0{m+x}}^{\tilde{\mathbb{O}}}-\gamma_{0m}^{\tilde{\mathbb{O}}}}{\beta_0}\nonumber\\
	&= \frac{24}{11}\left( \frac{2}{m}-\frac{2}{m+1}+\sum_{k = m+1}^{m+x}\frac{1}{k}-\frac{2}{m+x}+\frac{2}{m+x+1}\right)\nonumber\\
	&=\frac{24}{11}K_m(x)\,,
\end{align}
with
\begin{equation}
	K_m(x) = \frac{2}{m}-\frac{2}{m+1}+\sum_{k = m+1}^{m+x}\frac{1}{k}-\frac{2}{m+x}+\frac{2}{m+x+1}\,.
\end{equation}
Clearly, this sum matches the form in Eq. \eqref{eq:general}, with coefficients $c_k=\pm1,\pm2$ and no gaps. We again parametrize $x$ as
\begin{align}
	x = m+t\qquad t\geq0\,.
\end{align}
This leaves out a finite number of values of $x<m$. Therefore, for $x\geq m$ we write
{\small
	\begin{align}
		\Delta_m^{\tilde{\mathbb{O}}}(m+t)=\frac{24}{11}\left(\frac{2}{m}-\frac{2}{m+1}+\sum_{k = m+1}^{2m+t}\frac{1}{k}-\frac{2}{2m+t}+\frac{2}{2m+t+1}\right)\,.
	\end{align}
}
The generalized argument applies straightforwardly to the prime in the interval 
\begin{equation}
	m+1+\frac{t}{2}\leq p\leq 2m+t\,,
\end{equation}
so that $\Delta_m^{\tilde{\mathbb{O}}}(m+t)$ is not an integer. \par
We have  left out the values of $x< m$.
As before, we demonstrate that
\begin{align}
	\label{emtilde}
	K_m(x) <\log(2)<1,\qquad\qquad \forall x<m\,.
\end{align}	
Indeed,
\begin{equation}\label{eq:ineq3}
	K_m(x)\leq K_m( m-1) <	\lim_{m\rightarrow\infty}K_m( m-1)  = \log(2)
\end{equation}
because $K_m(x)$ is monotonically increasing in $x$ for fixed $m$ and $K_m(m-1)$ is monotonically increasing in $m$ by Lemma \ref{lemma3}.\par
The limit in Eq. \eqref{eq:ineq3}  is calculated by means of the Euler-Maclaurin formula \cite{EMformula}
\begin{equation}
	K_m( m-1) =\log(2)-\frac{3}{4 m}+\frac{25}{16 m^2}-\frac{9}{4 m^3}+O\left(\frac{1}{m^4}\right)\,.
\end{equation}
Hence, by Eq. \eqref{eq:ineq3}
\begin{align}
	\Delta_m^{\tilde{\mathbb{O}}}(x)= 	\frac{24}{11}K_m(x) <\frac{24}{11}\log(2)<1.53\,,\qquad\qquad\forall x<m\,.
\end{align}
Therefore, $\Delta_m^{\tilde{\mathbb{O}}}(x)$  is strictly lower than $2$.\\
We conclude that $\Delta_m^{\tilde{\mathbb{O}}}$ cannot be an integer for $x\geq m$ and cannot be an integer larger then $1$ for $x<m$, thus proving the theorem.
\qed
\section{Conclusions}
We have demonstrated that the eigenvalues of the (diagonal) matrices  $\frac{\gamma_0}{\beta_0}$ for the twist-$2$ operators in SU($N$) Yang-Mills theory satisfy the nonresonant condition in \textbf{Theorem \ref{th:1}}. Accordingly, the nonresonant diagonal scheme exists for all twist-$2$ operators in SU($N$) Yang-Mills theory, where the renormalized mixing matrices $Z(\lambda)$ in Eq. \eqref{ZZ} are one-loop exact with eigenvalues \cite{Bochicchio:2021geometry}
\begin{equation}
	Z_{\mathcal{O}_i}(\lambda) = \Bigg(\frac{g(\mu)}{g(\frac{\mu}{\lambda})}\Bigg)^{\frac{\gamma_{0\mathcal{O}_i}}{\beta_0}} \,.
\end{equation}
We conclude that the UV asymptotics of the generating functional of correlators calculated in \cite{BPSpaper2} applies to all the twist-$2$ operators in SU($N$) Yang-Mills theory. 
\section*{Acknowledgements}
I would like to thank Marco Bochicchio and Federico Pellarin for reading the manuscript.

\bibliographystyle{JHEP}
\bibliography{mybib} 

\providecommand{\href}[2]{#2}\begingroup\raggedright\begin{thebibliography}{10}

\bibitem{BPSpaper2}
M.~Bochicchio, M.~Papinutto and F.~Scardino, \emph{{UV asymptotics of n-point
  correlators of twist-2 operators in SU(N) Yang-Mills theory}},
  \href{http://dx.doi.org/10.1103/PhysRevD.108.054023}{\emph{Phys. Rev. D} {\bf
  108} (2023) 054023}, [\href{https://arxiv.org/abs/2208.14382}{{\tt
  2208.14382}}].

\bibitem{Bochicchio:2021geometry}
M.~Bochicchio, \emph{{On the geometry of operator mixing in massless QCD-like
  theories}},
  \href{http://dx.doi.org/10.1140/epjc/s10052-021-09543-5}{\emph{Eur. Phys. J.
  C} {\bf 81} (2021) 749}, [\href{https://arxiv.org/abs/2103.15527}{{\tt
  2103.15527}}].

\bibitem{BPSL}
M.~Bochicchio, M.~Papinutto and F.~Scardino, \emph{{On the structure of the
  large-$N$ expansion in SU($N$) Yang-Mills theory}},
  \href{https://arxiv.org/abs/2401.09312}{{\tt 2401.09312}}.

\bibitem{Becchetti:2021for}
M.~Becchetti and M.~Bochicchio, \emph{{Operator mixing in massless QCD-like
  theories and Poincar\`e\textendash{}Dulac theorem}},
  \href{http://dx.doi.org/10.1140/epjc/s10052-022-10551-2}{\emph{Eur. Phys. J.
  C} {\bf 82} (2022) 866}, [\href{https://arxiv.org/abs/2103.16220}{{\tt
  2103.16220}}].

\bibitem{Belitsky:1998gc}
A.~V. Belitsky and D.~Mueller, \emph{{Broken conformal invariance and spectrum
  of anomalous dimensions in QCD}},
  \href{http://dx.doi.org/10.1016/S0550-3213(98)00677-4}{\emph{Nucl. Phys. B}
  {\bf 537} (1999) 397--442}, [\href{https://arxiv.org/abs/hep-ph/9804379}{{\tt
  hep-ph/9804379}}].

\bibitem{Braun:2003rp}
V.~M. Braun, G.~P. Korchemsky and D.~Mueller, \emph{{The Uses of conformal
  symmetry in QCD}},
  \href{http://dx.doi.org/10.1016/S0146-6410(03)90004-4}{\emph{Prog. Part.
  Nucl. Phys.} {\bf 51} (2003) 311--398},
  [\href{https://arxiv.org/abs/hep-ph/0306057}{{\tt hep-ph/0306057}}].

\bibitem{Belitsky:2004sc}
A.~V. Belitsky, S.~E. Derkachov, G.~P. Korchemsky and A.~N. Manashov,
  \emph{{Dilatation operator in (super-)Yang-Mills theories on the
  light-cone}},
  \href{http://dx.doi.org/10.1016/j.nuclphysb.2004.11.034}{\emph{Nucl. Phys. B}
  {\bf 708} (2005) 115--193}, [\href{https://arxiv.org/abs/hep-th/0409120}{{\tt
  hep-th/0409120}}].

\bibitem{BPS1}
M.~Bochicchio, M.~Papinutto and F.~Scardino, \emph{{n-point correlators of
  twist-2 operators in SU(N) Yang-Mills theory to the lowest perturbative
  order}}, \href{http://dx.doi.org/10.1007/JHEP08(2021)142}{\emph{JHEP} {\bf
  08} (2021) 142}, [\href{https://arxiv.org/abs/2104.13163}{{\tt 2104.13163}}].

\bibitem{Beisert:2004fv}
N.~Beisert, G.~Ferretti, R.~Heise and K.~Zarembo, \emph{{One-loop QCD spin
  chain and its spectrum}},
  \href{http://dx.doi.org/10.1016/j.nuclphysb.2005.04.004}{\emph{Nucl. Phys. B}
  {\bf 717} (2005) 137--189}, [\href{https://arxiv.org/abs/hep-th/0412029}{{\tt
  hep-th/0412029}}].

\bibitem{ireland1990classical}
K.~Ireland and M.~I. Rosen, \emph{A classical introduction to modern number
  theory}, vol.~84.
\newblock Springer Science \& Business Media, 1990.

\bibitem{cebby}
P.~{\v{C}}eby{\v{s}}ev, \emph{M{\'e}moire sur les nombres premiers}.
\newblock Universit{\"a}tsbibliothek Johann Christian Senckenberg, 2010.

\bibitem{ramanujan1919proof}
S.~Ramanujan, \emph{A proof of bertrand’s postulate}, {\emph{Journal of the
  Indian Mathematical Society} {\bf 11} (1919) 27}.

\bibitem{EMformula}
T.~M. Apostol, \emph{An elementary view of euler's summation formula},
  {\emph{The American Mathematical Monthly} {\bf 106} (1999) 409--418}.

\bibitem{Negura}
J.~Nagura, \emph{{On the interval containing at least one prime number}},
  \href{http://dx.doi.org/10.3792/pja/1195570997}{\emph{Proceedings of the
  Japan Academy} {\bf 28} (1952) 177 -- 181}.

\end{thebibliography}\endgroup
	
\end{document}